\documentclass[a4paper,12pt]{article}
\pdfoutput=1

\usepackage{amssymb} 
\usepackage{amsmath}
\usepackage{graphicx}

\makeatletter

\@addtoreset{equation}{section}
\makeatother
\def\be{\begin{equation}}
\def\ee{\end{equation}}
\def\bea{\begin{eqnarray}}
\def\eea{\end{eqnarray}}

\def\({\left(}
\def\){\right)}
\def\<{\left<}
\def\>{\right>}

\def\be{\begin{equation}}
\def\ee{\end{equation}}
\def\bea{\begin{eqnarray*}}
\def\eea{\end{eqnarray*}}
\def\ben{\begin{eqnarray}}
\def\een{\end{eqnarray}}
\def\({\left(}
\def\){\right)}
\def\<{\left<}
\def\>{\right>}
\def\!{\right|}
\def\|{\left|}

\def\[{\left[}
\def\]{\right]}

\def\+{\bar}
\def\mb{\mathbb}

\def\L{{\cal{L}}}
\def\t{\widetilde}

\def\R{{\cal{R}}}

\def\N{{\cal{N}}}

\def\L{{\cal{L}}}

\def\eps{{\cal{\varepsilon}}}

\def\E{{\cal{E}}}

\begin{document}

\setlength{\unitlength}{1mm}

\pagestyle{empty}
\vskip-10pt
\vskip-10pt
\hfill %{\tt hep-th/yymmnnn}
\begin{center}
\vskip 3truecm
{\Large \bf
Lightlike conformal reduction\\
of 6d $(1,0)$ theories}
\vskip 2truecm
{\large \bf
Andreas Gustavsson}
\vspace{1cm} 
\begin{center} 
Physics Department, University of Seoul, Seoul 02504 KOREA
\end{center}
\vskip 0.7truecm
\begin{center}
(\tt agbrev@gmail.com)
\end{center}
\end{center}
\vskip 2truecm
{\abstract We study 6d $(1,0)$ superconformal theories. These have a natural lightlike conformal Killing vector, the Dirac current. We perform a conformal dimensional reduction along the Dirac current down to five-dimensions in such a way that we always preserve at least two real supercharges.}

\vfill
\vskip4pt
\eject
\pagestyle{plain}

\section{Introduction}
For the superconformal M5 brane in absence of supergravity background fields we have the 6d conformal Killing spinor equation 
\ben
\nabla_M \eps &=& \Gamma_M \eta\label{CKE}
\een
which restricts the six-dimensional Lorentzian worldvolume geometry of the M5 brane. Here $\Gamma_M$ are 6d gamma matrices, $\eps$ and $\eta$ are four-component spacetime spinors of opposite Weyl chiralities, and each transform as a four-component spinor under the $Sp(2) = SO(5)$ R-symmetry rotations. There is a conjectured duality between the M5 brane theory and a theory we get by dimensional reduction along a circle \cite{Lambert:2010iw}, \cite{Douglas:2010iu}. This suggests that we may study the M5 brane theory using a 5d theory, which has an ordinary gauge field and hence a nonabelian generalization. If $v^M$ is a spatial Killing vector that generates the circle, then supersymmetries that remain in 5d satisfy
\ben
\L_v \eps &=& 0\label{ADD}
\een
in addition to (\ref{CKE}). This additional equation (\ref{ADD}) may not be satisfied by any of the solutions to (\ref{CKE}) in which case all supersymmetries are broken by the dimensional reduction. If all supersymmetries are broken, it becomes difficult to check if the duality between the 6d and 5d theories still holds. Without supersymmetry it is hard to do quantum computations in the 5d theory, which is nonrenormalizable. On the other hand, the full 6d superconformal symmetry should be present in the 5d quantum theory if the duality between the 6d and 5d theories is correct. The full superconformal symmetry may not be a symmetry of the classical 5d Lagrangian. But we expect that it is a hidden symmetry in the quantum theory that emerges when we sum over all instanton sectors. 

For an M5 brane on $\mb{R}^{1,5}$ there is an $OSp(8|2)$ superconformal symmetry whose bosonic subalgebra is $SO(2,6) \times Sp(2)$. Here $SO(2,6)$ is the conformal symmetry and $Sp(2) = SO(5)$ is the R-symmetry. These bosonic symmetries correspond to isometries of $AdS_7 \times S^4$, which is the dual geometry in the AdS-CFT correspondence. By compactifying one spatial direction we get $\mb{R}^{1,4} \times S^1$. Upon dimensional reduction to 5d super-Yang-Mills on $\mb{R}^{1,4}$ the conformal symmetry $SO(2,6)$ is broken to the Poincare symmetry $ISO(1,4)$. The M5 brane on $\mb{R}^{1,5}$ has $32$ real supercharges, parametrized by $16$ real Poincare supersymmetry parameters $\eps_P$ and $16$ special conformal supersymmetry parameters $\eta$. The solutions to (\ref{CKE}) are given by $\eps = \eps_P + \Gamma_M \eta x^M$. Upon circle compactification, only the $16$ Poincare supercharges corresponding to $\eps_P$ survive. 

Inspired by the BLG theory \cite{Bagger:2007jr}, \cite{Gustavsson:2007vu} and the ABJM theory \cite{Aharony:2008ug} for M2 branes on $\mb{R}^{1,2}$ with an $SO(2,3)$ conformal symmetry and an $SO(8)$ and $SU(4)$ R-symmetry respectively, corresponding to isometries of $AdS_4 \times S^7$ and $AdS_4 \times CP^3$, we may for the M5 brane instead consider theories with symmetries of $AdS_7 \times S^4$ and $\t{CP}^3 \times S^4$ respectively. Here $\t{CP}^3$ has $SU(1,3)$ isometry and is a noncompact version of $CP^3$. To obtain the corresponding field theory we may start with M5 brane on $\mb{R}^{1,5}$ and perform a certain conformal transformation that makes a lightlike direction compact and dimensionally reduce along that lightlike circle to obtain a 5d super Yang-Mills with $SU(1,3)$ conformal symmetry. The embedding of $SU(1,3)$ in $SO(2,6)$ corresponds to the Hopf fibration of $AdS_7$ over $\t{{C}P}^3$. The dimensionally reduced super Yang-Mills preserves $24$ real supercharges and was studied from various viewpoints in \cite{Lambert:2019jwi}, \cite{Lambert:2019fne}, \cite{Lambert:2020zdc} and also generalized to (1,0) superconformal theories in \cite{Lambert:2020jjm}.

In the above two examples the geometry of the M5 brane was conformally flat. We may also consider other kinds of geometries, for example $\mb{R}^{1,1} \times $TaubNUT, that admit conformal Killing spinors. If for some geometry there are $n$ solutions to the 6d conformal Killing spinor equation (\ref{CKE}), then because of the hidden four-component R-symmetry spinor index, there are in total $4n$ real supercharges. In flat space we have $n= 4+4 = 8$. But we will now continue to refer to the case of $4n$ real supercharges on a curved spacetime as (2,0) supersymmetry. The smallest amount of supersymmetry we can have is (1,0) supersymmetry corresponding to $Sp(1) = SU(2)$ R-symmetry and $2n$ real supercharges. 

A complete classification of Lorentzian six-manifolds that admit conformal Killing spinors with $Sp(2)$ R-symmetry is still lacking. But for $Sp(1)$ R-symmetry, a complete classification of Lorentzian six-manifolds that have conformal Killing spinors is found in \cite{Baum}. The difficulty with extending this classification to $Sp(2)$ R-symmetry of the (2,0) theory, is related to the fact that the Dirac current $\bar\eps \Gamma^M \eps$ in (2,0) theory is not necessarily lightlike. It is lightlike only if the matrix $\bar\eps\Gamma^M \gamma^A \eps$ has rank one. Here $\gamma^A$ denote the gamma matrices of the $Sp(2) = SO(5)$ R-symmetry where the index ranges over $A = 1,2,3,4,5$. 

In this paper we will take the R-symmetry to be $Sp(1) = SU(2)$ where we have the 6d conformal Killing spinor equation
\ben
\nabla_M \eps_I &=& \Gamma_M \eta_I\label{CKE2}
\een
Again $\Gamma_M$ are 6d gamma matrices, and we display the $SU(2)$ R-symmetry indices $I,J,... = 1,2$ explicitly. We have an invariant antisymmetric tensor $\eps_{IJ}$ and its inverse $\eps^{IJ}$ that we, by a convention, use to rise and lower indices by acting from the right, $\psi^I = \psi_J \eps^{JI}$ and $\psi_I = \psi^J \eps_{JI}$. Here $\psi_I$ represents any field or parameter that carries the index $I$.

We will show that we may reduce a Lorentzian 6d $(1,0)$ superconformal theory with $2n$ real supercharges down to a 5d supersymmetric Yang-Mills theory while preserving at least two real supercharges, without imposing the extra condition (\ref{ADD}) on the supersymmetry parameter. We pick one pair of supersymmetry parameters $\eps_I$ for $I=1,2$, and construct the Dirac current 
\bea
V^M &=& \bar\eps^I \Gamma^M \eps_I
\eea
which will always be a lightlike conformal Killing vector \cite{Baum}, \cite{Bak:2024ihe}. We then perform a conformal reduction along $V^M$. 

If we pick another solution to (\ref{CKE2}) then we get another Dirac current and by conformal reduction we get another supersymmetric 5d theory. Therefore, in the most generic case, we may only preserve two real supercharges, namely those which correspond to our choice of two supersymmetry parameters $\eps_I$ for $I = 1,2$ that we use to define $V^M = \bar\eps^I \Gamma^M \eps_I$. 

Our conformal reduction is a generalization of an ordinary dimensional reduction to the case when the vector field along which we dimensionally reduce is a conformal Killing vector field. For an ordinary Killing vector field $v^M$ we have the isometry condition of the metric tensor $g_{MN}$, 
\bea
\L_v g_{MN} &=& 0
\eea
Under dimensionsional reduction along this isometry, we put 
\bea
\L_v \Phi &=& 0
\eea
for all the fields, here collectively denoted as $\Phi$. For our conformal Killing vector field $V^M = \bar\eps^I \Gamma^M \eps_I$, we instead have  
\bea
\L_V g_{MN} &=& \frac{\Omega}{3} g_{MN}
\eea
where $\Omega =  \nabla_P V^P = 12 \bar\eps^I \eta_I$. What stands on the right-hand side may be thought of as an infinitesimal Weyl transformation of the metric tensor if we make $\Omega$ infinitesimal. A general infinitesimal Weyl transformation of the metric may be written as
\bea
\delta g_{MN} &=& 2 \sigma g_{MN}
\eea
where $\sigma$ is an arbitrary infinitesimal scalar field parametrizing the Weyl transformation. The Weyl weight of the metric is $\Delta_g = 2$ so we may write
\bea
\L_V g_{MN} &=& \frac{\Delta_g \Omega}{6} g_{MN}
\eea
To perform dimensional reduction along a conformal Killing vector, a natural guess seems to be that we should constrain the fields, collectively denoted as $\Phi$, as
\ben
\L_V \Phi &=& \frac{\Delta_{\Phi} \Omega}{6}  \Phi\label{notsusyred}
\een
where $\Delta_{\Phi}$ is the Weyl weight of the field under consideration. That is
\bea
\delta \Phi &=& \Delta_{\Phi} \sigma \Phi
\eea
under an infinitesimal Weyl transformation. In order to preserve supersymmetry under the dimensional reduction, it turns out that we need to also add an R-rotation term. To motivate this R-rotation, we note that by a straightforward computation that can be found in an appendix in \cite{Bak:2024ihe}, we have 
\bea
\L_V \eps_I &=& \frac{\Omega}{12} \eps_I + 8 \eps_J R^J{}_I
\eea
where $R^J{}_I$ is a generator of $SU(2)$ R-symmetry that we define as the following traceless part
\bea
R^I{}_J &=& \bar\eps^I \eta_J - \frac{1}{12} \delta^I_J \Omega 
\eea
As we will show in this paper, a supersymmetry preserving dimensional reduction is achieved by modifying (\ref{notsusyred}) by adding an R-rotation term as 
\ben
\L_V \Phi_I &=& \frac{\Delta_{\Phi} \Omega}{6} \Phi_I + 8 \Phi_J R^J{}_I\label{susyred}
\een
for all fields that rotate under R-symmetry. Here we display this for a field $\Phi_I$ in the fundamental representation of $SU(2)$ with one index $I$ downstairs, but this generalizes straightforwardly to any field, or composite field, in any R-symmetry representation by taking tensor products of the fundamental representation. We notice that $\Omega$ changes under a Weyl transformation whereas $R^I{}_J$ is invariant. This may be seen by noting the Weyl transformation rules,
\bea
g_{MN} &\rightarrow & e^{2\sigma} g_{MN}\cr
\eps_I &\rightarrow & e^{\frac{\sigma}{2}} \eps_I\cr
\eta_I &\rightarrow & e^{-\frac{\sigma}{2}} \(\eta_I + \frac{1}{2} \Gamma^M \eps_I \partial_M \sigma\)
\eea
We may use these to show that 
\bea
V^M &\rightarrow & V^M\cr
V_M &\rightarrow & e^{2\sigma} V_M\cr
\Omega &\rightarrow & \Omega + 6 \L_V \sigma\cr
R^I{}_J &\rightarrow & R^I{}_J
\eea
These transformations show that we may make a Weyl transformation with transformation parameter $\sigma$ that satisfies
\bea
\L_V \sigma &=& - \frac{\Omega}{6}
\eea
such that a conformal Killing vector $V^M$ becomes an ordinary Killing vector with respect to the Weyl transformed metric. 

To make this discussion more explicit, we may choose a coordinate system where the metric takes the form
\bea
ds^2 &=& e^{-2\sigma(x^+)} \(dx^+ dx^- + ...\)
\eea
We may then perform a Weyl transformation that removes the conformal factor, 
\bea
\t{ds^2} &=& dx^+ dx^- + ...
\eea
Then we have a Killing vector $\t{V} = \frac{\partial}{\partial x^+}$ and hence we can make a periodic identification $x^+ \sim x^+ + 2 \pi R$ and that is true even if $\sigma(x^+)$ is not periodic. After the Weyl transformation, the supersymmetric dimensional reduction amounts to putting 
\ben
\partial_+ \t\Phi_I &=& 8 \t\Phi_J R^J{}_I\label{above}
\een
for all fields collectively denoted $\Phi_I$ (and as before, for brevity we just put one index $I$ downstairs, but this can be generalized). If we have the Weyl transformation rules
\bea
\t{g}_{MN} &=& e^{2\sigma(x^+)} g_{MN}\cr
\t\Phi_I &=& e^{\Delta_{\Phi}\sigma(x^+)} \Phi_I
\eea
of the metric and the field, then for the original metric, the vector field $V = \t{V} = \partial_+$ becomes a conformal Killing vector field, and (\ref{susyred}) becomes
\ben
\partial_+ \Phi_I &=& -\Delta_{\Phi} (\partial_+ \sigma) \Phi_I + 8 \Phi_J R^J{}_I\label{locale}
\een
which is easily shown to be equivalent with (\ref{above}) using the above Weyl transformation. By noting that $R^I{}_J$ is covariantly constant, and hence also constant as it carries no spacetime indices (for a derivation of this fact, we refer to an appendix in \cite{Bak:2024ihe}), we can integrate (\ref{locale}) around the circle to get the global identification
\bea
\Phi_I(x^+ + 2\pi R) &=& e^{- \Delta_{\Phi} \(\sigma(x^+ + 2\pi R) - \sigma(x^+)\)} \Phi_J(x^+) \R^J{}_I
\eea
where $\R^J{}_I = \exp 8 x^+ R^J{}_I$. If $R^J{}_I$ had not been constant along the circle, then we would have had a path-ordered exponent, but since $R^J{}_I$ indeed is constant that simplifies to an ordinary exponent. We thus find a periodic boundary condition that is twisted by both the Weyl symmetry and the R-symmetry. The R-symmetry is a global symmetry, but we can gauge this R-symmetry by coupling the theory to a background R-symmetry gauge field. Then for an R-gauge field that is locally pure gauge, we can gauge it to zero. But this may involve a gauge parameter that is not necessarily periodic and this gauge parameter gives rise to the R-symmetry twisted boundary condition \cite{Kim:2012ava}, \cite{Kallen:2012va}, \cite{Gustavsson:2015fra}. The Weyl symmetry is already a local symmetry, and hence it is a gauge symmetry, and therefore we may also allow for a Weyl symmetry twist in our boundary condition.

Ordinary (or untwisted) dimensional reduction along a lightlike direction amounts to a discrete light cone quantization, or DLCQ for short. If we apply DLCQ to the M5 brane theory, it is conjectured to reduce to a supersymmetric quantum mechanics on the moduli space of instantons \cite{Aharony:1997an}, \cite{Aharony:1997th}, \cite{Lambert:2011gb}. This has also been extended to 6d (1,0) superconformal theories \cite{Aharony:1997pm}. It would be interesting to see how our conformal dimensional reduction with an R-symmetry twist would modify this DLCQ proposal.

In section 2 we present the result, the Lagrangian and the supersymmetry variations. In section 3 we show that the conformal reduction commutes with the supersymmetry variations and hence it is a consistent truncation. In section 4 we show that the Lagrangian is invariant under the proposed supersymmetries. In the appendices we collect derivations of various geometric formulas.

\section{Lightlike conformal reduction}
Lightlike dimensionsal reduction of the abelian selfdual tensor field $H_{MNP}$ has been studied in \cite{Lambert:2020scy}, \cite{Gustavsson:2023zny}. We assume that we have two lightlike conformal Killing vectors $V^M$ and $U^M$ whose inner product is denoted $V^M U_M = \N$ and will be assumed to be everywhere nonzero (and hence we may choose $\N>0$ by convention). We define 
\bea
G_{MN} &=& H_{MNP} U^P\cr
F_{MN} &=& H_{MNP} V^P\cr
K_M &=& H_{MNP} U^N V^P
\eea
We invert these relations as
\bea
H_{MNP} &=& \t{H}_{MNP} + \frac{3}{\N} \t{F}_{MN} U_P + \frac{3}{\N} \t{G}_{MN} V_P - \frac{6}{\N^2} K_M U_N V_P
\eea
Here 
\bea
G_{MN} &=& \t{G}_{MN} - \frac{2}{\N} K_M U_N \cr
F_{MN} &=& \t{F}_{MN} + \frac{2}{\N} K_M V_N
\eea
where tildes indicate quantities whose contractions with $U^M$ and $V^M$ are zero. We might put a tilde on $K_M$ for free, since $K_M = \t{K}_M$. The antisymmetric tensor induces an antisymmetric tensor in the four transverse directions,
\bea
\E^{MNPQ} &=& \frac{1}{\N} \eps^{MNPQRS} V_R U_S
\eea
The selfduality $H_{MNP} = \frac{1}{6} \eps_{MNP}{}^{QRS} H_{QRS}$ leads to 
\bea
\t{F}_{MN} &=& \frac{1}{2} \E_{MN}{}^{PQ} \t{F}_{PQ}\cr
\t{G}_{MN} &=& - \frac{1}{2} \E_{MN}{}^{PQ} \t{G}_{PQ}
\eea
and 
\bea
\t{H}_{MNP} &=& \E_{MNP}{}^Q K_Q
\eea
that we use to eliminate $\t{H}_{MNP}$. Following \cite{Lambert:2020scy}, \cite{Gustavsson:2023zny} we take antiselfduality of $\t{G}_{MN}$ off-shell and then the selfduality of $\t{F}_{MN}$ follows from a Lagrangian where $\t{G}_{MN}$ acts as a Lagrange multiplier,
\bea
\L &=& \frac{1}{2\N} \t{G}^{MN} \t{F}_{MN} + ...
\eea
For a nonabelian gauge group we do not have this starting point at a selfdual $H_{MNP}$ but we can keep $\t{F}_{MN}$, $\t{G}_{MN}$ and $K_M$ that we promote to nonabelian fields and search for a Lagrangian that has $(1,0)$ superconformal symmetry. There is a vector multiplet with the bosonic fields the gauge potential $A_M$ and a real scalar $\phi$ and the fermionic fields are two fermions $\lambda_I$ subject to a symplectic Majorana condition. The gauge field and the scalar carry $3+1$ degrees of freedom, and the fermions carry $8$ degrees of freedom. There may depending on the formulation also be extra auxiliary fields, such as here for example the $\t{G}_{MN}$ that carry no physical degrees of freedom. To this vector multiplet we may couple an arbitrary number $r$ of hypermultiplets whose bosonic fields are $2r$ scalar fields $q^{AI}$ subject to a reality condition and $r$ fermions $\psi^A$, also subject to a reality condition. Here $A = 1,...,r$ labels the different hypermultiplets. There are $2r$ bosonic degrees of freedom and $4r$ fermionic degrees of freedom. For convenience we will assume that the hypermultiplet fields are in the adjoint represenation of the gauge group. The Lagrangian is given by the sum $\L = \L_A + \L_{CS} + \L_m^{tensor} + \L_m^{hyper}$ where 
\bea
\L_A &=& \frac{1}{2 \N} \t{G}^{MN} \t{F}_{MN} + \frac{1}{2\N^2} K^M K_M\cr
\L_{CS} &=& - \frac{1}{4\N} \eps^{MNPQRS} \omega(A)_{MNP} \Omega_{QR} U_S
\eea
and
\bea
\L_m^{tensor} &=& - \frac{1}{2} (D_M \phi)^2 - \frac{R}{10} \phi^2\cr
&& + \frac{i}{2} \bar\lambda^I \Gamma^M D_M \lambda_I - \frac{e}{2} \bar\lambda^I \Gamma_M [\lambda_I,\phi] V^M
\eea
\bea
\L_m^{hyper} &=& - \frac{1}{4} D_M q^{AI} D_M q_{IA} - \frac{R}{20} q^{AI} q_{IA}\cr
&& + \frac{i}{2} \bar\psi_A \Gamma^M D_M \psi^A \cr
&& + \frac{e}{2} \bar\psi_A \Gamma_M  [\psi^A,\phi] V^M + e \bar\lambda^I \Gamma_M [\psi^A,q_{IA}] V^M
\eea
Here we define 
\ben
\Omega_{MN} &=& \partial_M \(\frac{U_N}{\N}\) - \partial_N \(\frac{U_M}{\N}\)\label{O}
\een
which satisfies 
\bea
\Omega_{MN} V^N &=& 0\cr
\Omega_{MN} U^N &=& 0
\eea
We show these properties Appendix \ref{Omegas}. The normalization of the Chern-Simons term $\L_{CS}$ is such that its infinitesimal variation is given by $\delta \omega(A)_{MNP} = \delta A_{[M} F_{NP]}$. An explicit expression is 
\bea
\omega(A)_{MNP} &=& A_{[M} \partial_N A_{P]} - \frac{2i}{3} A_{[M} A_N A_{P]}
\eea

We will show that this Lagrangian is invariant under
\bea
\delta \phi &=& - i \bar\eps^I \lambda_I\cr
\delta A_M &=& i \bar\eps^I \Gamma_{MN} \lambda_I V^N\cr
\delta \lambda_I &=&  \frac{1}{4} \Gamma^{MNP} \eps \(- \t{G}_{MN} \frac{V_P}{\N} + \t{F}_{MN} \frac{U_P}{\N} - 4 K_M \frac{U_N V_P}{\N^2}\)\cr
&& + \Gamma^M \eps_I D_M \phi + 4 \eta_I \phi\cr
\delta G_{MN} &=& - \frac{i}{2} D_Q \(\bar\eps^I \Gamma^Q \Gamma_{MNP} \lambda_I\) U^P\cr
\delta K_M &=& \(D_M \delta A_N - D_N \delta A_M\) U^N
\eea
and
\bea
\delta q^{AI} &=& 2 i \bar\eps^I \psi^A\cr
\delta \psi^A &=& - \Gamma^M \eps_I D_M q^{AI} - 4 \eta_I q^{AI} 
\eea
where all fields are restricted by the conformal reduction (\ref{Conf1}) and (\ref{Conf2}). There is a coupling constant $e$ that appears in various commutator terms and it appears in the gauge covariant derivatives of matter fields (collectively denoted by $\Phi$) as
\bea
D_M \Phi &=& \nabla_M \Phi - i e [A_M,\Phi]
\eea
For the purpose of showing that the classical Lagrangian is supersymmetric, the value of $e$ is not essential. Its value becomes important only in the quantum theory. 

The supersymmetry parameters $\eps_I$ satisfy the conformal Killing spinor equation 
\bea
\nabla_M \eps_I &=& \Gamma_M \eta_I
\eea

The Dirac current 
\bea
V^M &=& \bar\eps^I \Gamma^M \eps_I
\eea
is a lightlike conformal Killing vector. For $V^M$ to be nonzero we shall take $\eps_I$ to be commuting spinors. We have the following result, 
\ben
\L_V \eps_I &=& \frac{\Omega}{12} \eps_I + 8 \eps_J R^J{}_I\label{LVE}
\een
where we put 
\bea
\Omega &=& \nabla_M V^M\cr
R^I{}_J &=& \bar\eps^I \eta_J - \frac{1}{2} \delta^I_J \bar\eps^K \eta_K
\eea
The derivation of (\ref{LVE}) can be found in the appendix in \cite{Bak:2024ihe}. 

The Dirac conjugate is defined as
\bea
\bar\eps^I &=& (\eps_I)^* \Gamma^0\cr
\bar\eta^I &=& (\eta_I)^* \Gamma^0
\eea
where 
\bea
\Gamma \eps_I &=& - \eps_I\cr
\Gamma \eta_I &=& \eta_I
\eea
In flat Minkowski spacetime the chirality matrix is defined as $\Gamma = \Gamma^{012345}$, and it is numerically the same in curved spacetime. The symplectic Majorana condition reads 
\bea
\bar\eps^I &=& \eps_J^T C \eps^{JI}\cr
\bar\eta^I &=& \eta_J^T C \eps^{JI}
\eea
The 6d charge conjugation matrix $C$ has the properties 
\bea
C^T &=& C\cr
(C \Gamma^M)^T &=& - C \Gamma^M\cr
\eea
These relations can be used to show that $(\bar\eps_I \eta_J)^* = - \bar\eps^I \eta^J$. Since $R_{IJ} = R_{JI}$ (as a direct consequence of $R^I{}_I = 0$) we can see that $(R^J{}_I)^* = - R^I{}_J$ and hence this is an anti-hermitian generator of $SU(2)$. 

If we assume that $\Omega = 0$ then we must have $\eta_I = 0$ and then $V^M$ becomes a Killing vector and (\ref{LVE}) becomes $\L_V \eps_I = 0$ in which case this simply says that those supersymmetries survive under an ordinary lightlike dimensional reduction along $V^M$. 

We now ask ourselves if this familiar situation can be generalized to the case when $\Omega$ is not zero. As we discussed in the Introduction, we shall impose (\ref{susyred}) on all the fields, properly interpreted for fields in any representation of the R-symmetry. For the fields in the $(1,0)$ theory, we have $\Phi = (A_M,\phi,\lambda_I,\t{G}_{MN},K_M,q^{AI},\psi^A)$. Then for the tensor multiplet fields, we shall impose
\ben
\L_V \phi &=& - \frac{\Omega}{3} \phi\cr
\L_V \lambda_I &=& - \frac{5\Omega}{12} \lambda_I + 8 \lambda_J R^J{}_I\cr
\L_V A_M &=& 0\cr
\L_V \t{G}_{MN} &=& 0\cr
\L_V K_M &=& 0\label{Conf1}
\een
and for the hypermultiplet fields, we shall impose
\ben
\L_V q^{AI} &=& - \frac{\Omega}{3} q^{AI} - 8 R^I{}_J q^{AJ}\cr
\L_V \psi^A &=& - \frac{5\Omega}{12} \psi^A\label{Conf2}
\een
As we showed in  \cite{Bak:2024ihe} once we have $V^M$ and $(1,0)$ supersymmetry, we necessarily will also have another lightlike conformal Killing vector $U^M$. In order to show that the Lagrangian is superconformal, it is necessary to assume that their Lie derivatives commute,
\bea
[\L_V,\L_U] &=& 0
\eea

\section{Commuting Lie derivatives with supersymmetries}
We begin by showing that the supersymmetry variations commute with the conformal reduction by showing that 
\bea
\delta \L_V \Phi &=& \L_V \delta \Phi
\eea
on all fields $\Phi$ in the tensor and hypermultiplets. This shows that the conformal reduction is consistent with the supersymmetry variations.

\subsection{Checking $\delta q^{AI} = ...$}
We begin by checking if the conformal reduction is consistent with the variation
\bea
\delta q^{AI} &=& 2 i \bar\eps^I \psi^A
\eea
We act on both sides with $\L_V$, which yields
\bea
\L_V \delta q^{AI} &=& 2 i  \(\frac{1/2}{6} \Omega \bar\eps^I - 8 R^I{}_J \bar\eps^J\) \psi^A - 2 i \frac{5/2}{6} \Omega \bar\eps^I \psi^A\cr
\delta \L_V q^{AI} &=& - \frac{2}{6} \Omega \delta q^{AI} - 8 R^I{}_J \delta q^{AJ} 
\eea
and we see that indeed
\bea
\L_V \delta q^{AI} &=& \delta \L_V q^{AI}
\eea

\subsection{Checking $\delta \psi^A = ...$}
We now check the variation
\bea
\delta \psi^A &=& - \Gamma^M \eps_I \partial_M q^{AI} - 4 \eta_I q^{AI}
\eea
First we compute
\bea
\delta \L_V \psi^A &=& \delta \(- \frac{5\Omega}{12} \psi^A\)\cr
&=& - \frac{5\Omega}{12} \delta \psi^A\cr
&=& \frac{5\Omega}{12} \(\Gamma^M \eps_I D_M q^{AI} + 4 \eta_I q^{AI}\)
\eea
Next we shall compute 
\bea
\L_V \delta \psi^A &=& \L_V \(-\Gamma^M \eps_I D_M q^{AI}\) + \L_V \(- 4 \eta_I q^{AI}\) 
\eea
We begin by computing the Lie derivative acting the first term, which is a spinorial quantity (since the vector and R indices are fully contracted), so we use the usual formula for a Lie derivative acting on a spinorial quantity. Moreover, by $V^M A_M = 0$, we can use a gauge covariant derivative in the Lie derivative for free. Thus we compute
\bea
\L_V \(- \Gamma^M \eps_I D_M q^{AI}\) &=& - \(V^P D_P + \frac{1}{4} \nabla_P V_Q \Gamma^{PQ}\) \Gamma^M \eps_I D_M q^{AI}\cr
&=& - \Gamma^M \(\L_V \eps_I\) D_M q^{AI} - \Gamma^M \eps_I V^P D_P D_M q^{AI}\cr
&& - \frac{1}{2} \(\nabla_P V_M - \nabla_M V_P\) \Gamma^P \eps_I D^M q^{AI}
\eea
Now we look at the term
\bea
 - \Gamma^M \eps_I V^P D_P D_M q^{AI} &=& - \Gamma^M \eps_I D_M \(\L_V q^{AI}\) + \Gamma^M \eps_I (\nabla_M V_P) \nabla^P q^{AI}
\eea
The second term combines with the last term above, switching $\nabla_P V_M - \nabla_M V_P$ into $- \nabla_P V_M - \nabla_M V_P = - \frac{\Omega}{3} g_{PM}$. We get
\bea
\L_V \(- \Gamma^M \eps_I D_M q^{AI}\) &=&  - \Gamma^M \(\L_V \eps_I\) D_M q^{AI} - \Gamma^M \eps_I \nabla_M \(\L_V q^{AI}\)\cr
&& + \frac{\Omega}{6} \Gamma^M \eps_I \partial_M q^{AI}
\eea
Now we apply the conformal reduction, and then we find that the Weyl weights add up to $
- \frac{1}{12} + \frac{1}{3} + \frac{1}{6} = \frac{5}{12}$ and we get
\bea
\L_V \(- \Gamma^M \eps_I \partial_M q^{AI}\) &=& \frac{5\Omega}{12} \Gamma^M \eps_I D_M q^{AI} + \frac{\nabla_M \Omega}{3} \Gamma^M \eps_I q^{AI}
\eea
In the process we have noted that $R^I{}_J$ is covariantly constant \cite{Bak:2024ihe}.

Now we shall add the Lie derivative acting on the second term,
\bea
\L_V \(- 4 \eta_I q^{AI}\) &=& - 4 (\L_V \eta_I) q^{AI} - 4 \eta \L_V q^{AI}
\eea
In the Appendix we show that
\ben
\L_V \eta_I &=& - \frac{\Omega}{12} \eta_I + 8 \eta_J R^J{}_I + \frac{1}{12} \Gamma^M \eps_I \partial_M \Omega\label{strange}
\een
By applying conformal reduction on $\L_V q^{AI} = - \frac{\Omega}{3} q^{AI} - 8 R^I{}_J q^{AJ}$ and by noting that $\L_V F_{MN} = 0$, we get
\bea
\L_V \(- 4 \eta_I q^{AI}\) &=& 4 \frac{5\Omega}{12} \eta_I q^{AI} - \frac{\nabla_M \Omega}{3} \Gamma^M \eps_I q^{AI}
\eea
So by summing the two terms, we get
\bea
\L_V \delta \psi^A &=& \frac{5\Omega}{12} \Gamma^M \eps_I D_M q^{AI} + 4 \frac{5\Omega}{12} \eta_I q^{AI} 
\eea
Hence 
\bea
\L_V \delta \psi^A &=& \delta \L_V \psi^A
\eea

\subsection{Checking $\delta G_{MN} = ...$}
We first compute
\bea
\delta \L_V G_{MN} &=& 0
\eea
We shall next compute
\bea
\L_V \delta G_{MN} &=& ?
\eea
Let us first study the supersymmetry variation itself,
\bea
\delta G_{MN} &=& - \frac{i}{2} \(D_Q T^Q_{MNP}\) U^P\cr
T^Q_{MNP} &=& \bar\eps^I \Gamma^Q \Gamma_{MNP} \lambda_I
\eea
We find that
\bea
T_{Q,MNP} &=& 3 \(T_{[MN} g_{P]Q} + \frac{1}{6} \eps_{MNPQRS} T^{RS}\)
\eea
where
\bea
T_{MN} &=& \bar\eps^I \Gamma_{MN} \lambda_I
\eea
Since the gauge potential is a conformal weight zero, we will replace gauge covariant derivatives with geometric covariant derivatives, or alternatively consider abelian gauge group. The generalization to nonabelian case is then obtained by simply replacing back the gauge covariant derivatives, but there would be no changes in the computation that follows other than notational, since we have $\L_V A_M = 0$ and $V^M F_{MN} = 0$ under the conformal reduction. We now get
\bea
\nabla^Q T_{Q,MNP} &=& 3 \(\nabla_{[P} T_{MN]} + \frac{1}{6} \eps_{MNPQRS} \nabla^Q T^{RS}\)
\eea
so 
\bea
\delta G_{MN} &=& - \frac{i}{2} 3 \(\nabla_{[P} T_{MN]} + \frac{1}{6} \eps_{MNPQRS} T^{RS}\) U^P\cr
&=& - i \(U^P \nabla_P T_{MN} + 2 U^P \nabla_{[M} T_{N]P}\)\cr
&& + \frac{i}{2} 3 \(\nabla_{[P} T_{MN]} - \frac{1}{6} \eps_{MNPQRS} \nabla^Q T^{RS}\) U^P 
\eea
We may discard the last line from the variation because selfduality will be preserved when acting on both sides with $\L_V$ trivially. Basically $T_{MN} = \delta B_{MN}$ and the last line is nothing but $\delta H_{MNP}^- = 0$ which shall be respected by acting on it with $\L_V$. So we descend to 
\bea
\delta G_{MN} &=&  - i \(U^P \nabla_P T_{MN} + \nabla_M U^P T_{PN} + \nabla_N U^P T_{MP}\) \cr
&& + i \(\nabla_M \(U^P T_{PN}\) - \nabla_N \(U^P T_{PM}\)\)
\eea
That is,
\bea
\delta G_{MN} &=& - i \L_U T_{MN}\cr
&& + i \(\nabla_M \(U^P T_{PN}\) - \nabla_N \(U^P T_{PM}\)\)
\eea
Now we act on both sides with $\L_V$. Then 
\bea
\L_V \delta G_{MN} &=& - i \L_V \L_U T_{MN}\cr
&& + i \L_V \(\nabla_M \(U^P T_{PN}\) - \nabla_N \(U^P T_{PM}\)\)
\eea
The first line is zero by the fact that $\L_V T_{MN} = 0$ and $[\L_V,\L_U] = 0$. Let us put 
\bea
X_{MN} &=& \nabla_M T_N - \nabla_N T_M\cr
T_N &=& U^P T_{PN}
\eea
Then
\bea
\L_V X_{MN} = V^Q \nabla_Q X_{MN} + \nabla_M V^Q X_{QN} + \nabla_N V^Q X_{MQ}
\eea
After some cancelations, we get
\bea
\L_V X_{MN} &=& V^Q [\nabla_Q,\nabla_M] T_N \cr
&& +  \nabla_M \(V^Q \nabla_Q T_N + \nabla_N V^Q T_Q\)\cr
&&  - T_Q \nabla_M \nabla_N V^Q\cr
&& - V^Q [\nabla_Q,\nabla_N] T_M \cr
&& - \nabla_N \(V^Q \nabla_Q T_M + \nabla_M V^Q T_Q\)\cr
&& + T_Q \nabla_N \nabla_M V^Q 
\eea
Now we use that $\L_V T_M = 0$ and get
\bea
\L_V X_{MN} &=& V^Q [\nabla_Q,\nabla_M] T_N - V^Q [\nabla_Q,\nabla_N] T_M \cr
&&  - T_Q [\nabla_M,\nabla_N] V^Q
\eea
which we can write as
\bea
\L_V X_{MN} &=& V^Q T^S \(R_{QMNS} + R_{NQMS} + R_{MNQS}\) \cr
&=& 0
\eea
We conclude that 
\bea
\L_V \delta G_{MN} &=& 0
\eea
and hence
\bea
\delta \L_V G_{MN} &=& \L_V \delta G_{MM}
\eea

\subsection{Checking $\delta \phi = ...$}
First we compute
\bea
\delta \L_V \phi &=& - \frac{\Omega}{3} \delta \phi\cr
&=& \frac{i \Omega}{3} \bar\eps^I \lambda_I
\eea
Next we compute
\bea
\L_V \delta \phi &=& \L_V \(- i \bar\eps^I \lambda_I\)\cr
&=& - i \(\L_V \bar\eps^I\) \lambda_I - i \bar\eps^I \L_V \lambda_I
\eea
We now apply conformal reduction, and find that the Weyl weights add up as $\frac{1}{2} - \frac{5}{2} = - 2$ and we get
\bea
\L_V \delta \phi &=& \frac{i\Omega}{3} \bar\eps^I \lambda_I
\eea
and hence
\bea
\delta \L_V \phi &=& \L_V \delta \phi
\eea

\subsection{Checking $\delta A_M = ...$}
First we compute 
\bea
\delta \L_V A_M &=& 0
\eea
Next we compute 
\bea
\L_V \delta A_M &=& \L_V \(i \bar\eps^I \Gamma_{MN} \lambda_I V^N\)
\eea
Now we use $\L_V e_M^A = \frac{\Omega}{6} e_M^A$ for the vielbein and $\L_V V^M = 0$ together with conformal reduction of $\lambda_I$ and $\bar\eps^I$. This leads to the following numerology $\frac{1}{12} + \frac{2}{6} - \frac{5}{12} = 0$ and we get
\bea
\L_V \delta A_M &=& 0
\eea
and hence
\bea
\delta \L_V A_M &=& \L_V \delta A_M
\eea

\subsection{Checking $\delta \lambda_I = ...$}
First we compute
\bea
\delta \L_V \lambda_I &=& - \frac{5\Omega}{12} \delta \lambda_I + 8 \delta \lambda_J R^J{}_I
\eea
Next we compute
\bea
\L_V \delta \lambda_I &=& - \frac{1}{4} \L_V \(\Gamma^{MNP} \eps_I \t{G}_{MN} \frac{V_P}{\N} + ...\)\cr
&& + \L_V \(\Gamma^M \eps_I D_M \phi + 4 \eta_I \phi\) 
\eea
The computation of the second line proceeds in an analogous manner as we did the computation for $\L_V \delta \psi^A$. The first line is computed using $\L_V \Gamma^{MNP} = - \frac{\Omega}{2} \Gamma^{MNP}$ and $\L_V \eps_I = \frac{\Omega}{12} \eps_I + 8 \eps_J R^J{}_I$. The Weyl weights add up to $\frac{1}{2} - 3 = - \frac{5}{2}$ In summary, we get
\bea
\L_V \delta \lambda_I &=&  - \frac{5\Omega}{12} \delta \lambda_I + 8 \delta \lambda_J R^J{}_I
\eea
and hence 
\bea
\delta \L_V \lambda_I &=& \L_V \delta \lambda_I
\eea
It is also interesting to note in this context that $U_M/\N$ and $V_M/\N$ are conformal singlets, as we show in Appendix A.

\section{The superconformal Lagrangian} 
When showing that the Lagrangian is superconformal we may follow the computation in the Appendix of \cite{Gustavsson:2023zny}. Below we will highlight what changes in that computation as we change from ordinary lightlike to conformal lightlike Killing vectors. 

We get
\bea
\delta \L_A &=& - i \bar\eps^I \Gamma_{NP} \lambda_I V^P D_M \(\frac{1}{\N} \t{G}^{MN}\)\cr
&& + \frac{i}{4} \bar\eps^I \Gamma^Q \Gamma^{MNP} \lambda_I D_Q \(\t{F}_{MN} \frac{U_P}{\N}\)\cr
&& - i \bar\eps^I \Gamma_{NP} \lambda_I V^P D_M \(\frac{2}{\N^2} K^{[M} U^{N]}\)
\eea
and collecting all terms that involve $\t{G}_{MN}$ coming from the variation of the matter part we find 
\bea
\delta \L_m|_{\t{G}_{MN}} &=& i \bar\eps^I \Gamma_{NP} \lambda_I V^P D_M \(\frac{1}{\N} \t{G}^{MN}\) \cr
&& + \frac{i}{2} \bar\eps^I \Gamma^{MN} \lambda_I \(\L_V \(\frac{\t{G}_{MN}}{\N}\) +  \frac{\Omega}{3} \frac{\t{G}_{MN}}{\N}\)
\eea
The last lines vanishes when imposing the conformal reduction condition
\bea
\L_V \(\frac{\t{G}_{MN}}{\N}\) &=& - \frac{\Omega}{3} \frac{\t{G}_{MN}}{\N}
\eea
Next collecting terms involving $\t{F}_{MN}$ from the matter part, we find 
\bea
\delta \L_m|_{\t{F}_{MN}} &=& - \frac{i}{4} \bar\eps^I \Gamma^{MNP} \Gamma^Q \lambda_I D_Q \(\t{F}_{MN} \frac{U_P}{\N}\)
\eea
Adding the corresponding term in $\delta \L_A$ they combine into 
\bea
\frac{i}{4} \bar\eps^I \Gamma^{QMNP} \lambda_I \t{F}_{MN} \Omega_{PQ}
\eea
That we cancel by the Chern-Simons term. At last we collect terms involving $K_M$ from the matter part,
\bea
\delta \L_m|_{K_M} &=& - i \bar\lambda^I \Gamma^Q \Gamma^{MNP} \eps_I D_Q \(K_M \frac{U_N V_P}{\N^2}\)
\eea
Adding the corresponding contribution from $\delta \L_A$ we get by summing them up
\bea
i \bar\eps^I \Gamma_{NP} \lambda_I \(\L_V\(\frac{1}{\N^2}K_N U_P\) + \frac{\Omega}{3\N^2} K_N U_P\)
\eea
which is zero by the conformal reduction condition
\bea
\L_V\(\frac{1}{\N^2}K_N U_P\) &=& - \frac{\Omega}{3\N^2} K_N U_P
\eea
After cancelling most remaining terms in a standard manner, what remains then are two contributions that we will write as $\delta \L = \delta \L_m^{tensor} + \delta \L_m^{hyper}$. We will now look at each of these contributions in turn. First  
\bea
\delta \L_m^{tensor} &=& e \bar\lambda^I \Gamma^M \Gamma^N \eps_I V_M [D_N \phi,\phi] \cr
&& - e \bar\lambda^I \Gamma^{MN} \eps_I V_M [D_N \phi,\phi]
\eea
where the second line comes from varying $A_M$ in the scalar field kinetic term. Simplifying we get
\bea
\delta \L_m^{tensor} &=& 2 e \bar\lambda^I \eps_I [V^M D_M \phi,\phi] 
\eea
We note $V^M A_M = 0$, so what we find inside the commutator is $\L_V \phi = - \frac{\Omega}{3} \phi$ and hence the commutator is vanishing, so we get
\bea
\delta \L_m^{tensor} &=& 0
\eea
Next
\bea
\delta \L_m^{hyper} &=& - 2 e \bar\lambda^I \eps_J [V^M D_M q^{AJ},q_{IA}] - 4e \bar\lambda^I \Gamma_M \eta_J V^M [q^{AJ},q_{IA}]
\eea
The first term is rewritten using $\L_V q^{AJ} = - \frac{\Omega}{3} q^{AJ} - 8 R^J{}_K q^{AK}$ as
\bea
- 2 e \bar\lambda^I \eps_J [V^M D_M q^{AJ},q_{IA}] &=& \frac{2 e \Omega}{3} \bar\lambda^I \eps_J [q^{AJ},q_{IA}] + 16 e \bar\lambda^I \eps_J R^J{}_K [q^{AK},q_{IA}]
\eea
The second term is rewritten as using
\ben
V^M \nabla_M \eps_I &=& \frac{\Omega}{6} \eps_I + 4 \eps_J R^J{}_I\label{p}
\een
(that we derive in Appendix \ref{PP}) as
\bea
- 4e \bar\lambda^I \Gamma_M \eta_J V^M [q^{AJ},q_{IA}] &=& - 4e \bar\lambda^I V^M \nabla_M \eps_J [q^{AJ},q_{IA}]\cr
&=& - \frac{2 e \Omega}{3} \bar\lambda^I \eps_J [q^{AJ},q_{IA}] - 16 e \bar\lambda^I \eps_K R^K{}_J [q^{AJ},q_{IA}]
\eea
Adding up the contributions, we find 
\bea
\delta \L_m^{hyper} &=& 0
\eea
In summary then, we have shown that the Lagrangian is superconformal
\bea
\delta \L &=& 0
\eea
up to boundary terms. Let us now discuss these boundary terms. When we check the supersymmetry of the Lagrangian we performed integrations by parts to bring derivatives acting on fermions to instead act on bosonic fields, as our convention. In the process there appears boundary terms. These are on the form
\bea
\delta \L &=& D_M K^M
\eea
where
\bea
K^M &=& V^M K + U^M L + \t{K}^M
\eea
Then we notice that
\bea
D_M \(V^M K\) &=& V^M D_M K + \Omega K\cr
&=& \L_V K + \Omega K
\eea
Since $K$ has Weyl scaling dimension $-6$ we find that 
\bea
\L_V K &=& - \Omega K
\eea
The upshot is that $D_M \(V^M K\) = 0$ and so
\bea
D_M K^M &=& D_M \(U^M L + \t{K}^M\) 
\eea
reduces to boundary term of a five-dimensional theory as the component of the original $K^M$ along $V^M$ does not give any contribution.

\section{Singularities and boundaries}
Our discussion so far has been concerned with the local aspects of the dimensional reduction. In the end we need to integrate the Lagrangian density over the six-manifold, where we constrain the Lie derivative of the fields along the conformal vector field $V$ as
\ben
\L_V \Phi_I &=& \frac{\Delta_{\Phi} \Omega}{6} \Phi_I + 8 \Phi_J R^J{}_I\label{CE}
\een
We may encounter a singularity if the vector field $V$ vanishes on a submanifold. To see how such a singularity may arise explicitly, we will study a conformally flat spacetime with the metric  
\ben
ds^2 &=& e^{2\sigma} \(2 dx^+ dx^- + dx^i dx^i\)\label{CF}
\een
for $i = 1,2,3,4$ and where $\sigma$ can be an arbitrary function of all six coordinates. Then the solutions to the conformal Killing spinor equation are given by
\bea
\eps_I &=& e^{\frac{\sigma}{2}} \(\eps_{I0} + \Gamma_M \eta_I x^M\)
\eea
where $\eps_{I0}$ and $\eta_I$ are constant bosonic spinors that correspond to the Poincare and the special conformal supercharges respectively. The Dirac current is given by
\bea
V^M &=& \bar\eps^I \Gamma^M_g \eps_I
\eea
Here we define $\Gamma_g^M = e^{-\sigma} \Gamma^M$ where $\Gamma^M$ are the Minkowski space gamma matrices. If we expand the Dirac current in components, then we get
\bea
V^M &=& \bar\eps_0^I \Gamma^M \eps_0 + 2 \bar\eps_0^I \eta_I x^M + 2 \bar\eps_0^I\Gamma^{MN} \eta_I x_N\cr
&& + \bar\eta^I \Gamma^N \eta_I \(\delta^M_N |x|^2 - 2 x_N x^M\)
\eea 
Here we define $|x|^2 = 2 x^+ x^- + x^i x^i$. From this expansion we can see that the terms that appear in $V = V^M \partial_M$ correspond to a translation, a dilatation, a Lorentz transformation and a special conformal transformation, respectively. We may also expand 
\bea
\Omega = \nabla_M V^M = 12 \bar\eps_0^I \eta_I - 12 \bar\eta^I \Gamma_M \eta_I x^M
\eea
One may now explicitly see that $V^M$ does not depend on $\sigma$ whereas $\Omega$ depends on $\sigma$, which reflects the general fact that $V^M$ is Weyl invariant while $\Omega$ is not. If for a given $\Omega$ as computed with the flat Minkowski metric we can solve the equation
\ben
\L_V \sigma &=& - \frac{\Omega}{6}\label{eqOc}
\een
with respect to $\sigma$ then $V$ becomes a Killing vector in the metric (\ref{CF}) with this particular solution $\sigma$. 

First let us select two Poincare supercharges corresponding to $\eps_I = \eps_{I0}$ that we want to preserve under the dimensional reduction. Then $\Omega = 0$ and the Dirac current $V^M$ is a Killing vector. If we choose the two supercharges such that $V^M = \delta^M_+$ then we can perform the dimensional reduction by imposing the constraints that $\L_V = \partial_+$ vanishes on all fields. For the 6d tensor multiplet scalar field we start with the action
\bea
- \frac{1}{2} \int dx^+ dx^- d^4 x \(- 2 \partial_+ \phi \partial_- \phi + \partial_i \phi \partial_i \phi\)
\eea
By periodically identifying $x^+ \sim x^+ + 1$ and imposing the constraint $\partial_+ \phi = 0$ we obtain the dimensionally reduced action
\ben
- \frac{1}{2} \int dx^- d^4 x \partial_i \phi \partial_i \phi\label{ex1}
\een
If we assume that the metric of the 5d theory is $ds^2 = - c^2 (dx^-)^2 + dx^i dx^i$, then the corresponding action would have been
\bea
\frac{1}{2} \int dx^- d^4 x \(\frac{1}{c^2} (\partial_- \phi)^2 - \partial_i \phi \partial_i \phi\)
\eea
We recover our action (\ref{ex1}) in the nonrelativistic limit $c \rightarrow \infty$. The $(2,0)$ supersymmetric extension of (\ref{ex1}) was obtained in \cite{Lambert:2018lgt}, \cite{Gustavsson:2023zny}.

Let us next select two special conformal supercharges $\eps_I = \Gamma_M \eta_I x^M$ that we want to preserve under the dimensional reduction. If we assume that $\bar\eta^I \Gamma^M \eta_I = \delta^M_+$, then the conformal Killing vector has the components 
\bea
V^+ &=& r^2\cr
V^- &=& - 2 (x^-)^2\cr
V^i &=& - 2 x^- x^i
\eea
where $r^2 = x^i x^i$. This vector field vanishes along the lightlike line parametrized by $x^+$ where $x^- = 0$ and $r = 0$ and our aim is to explore what implications that will have for the dimensionally reduced theory. Let us examine the scalar field $\phi$ in the $(1,0)$ tensor multiplet whose scaling dimension is $\Delta_{\phi} = - 2$. The constraint equation for $\phi$ is given by
\bea
\L_V \phi &=& - \frac{\Omega}{3} \phi
\eea
Explicitly it becomes
\ben
r^2 \partial_+ \phi - 2 (x^-)^2 \partial_- \phi - 2 x^- x^i \partial_i \phi &=& 4 x^- \phi\label{C1}
\een
or, if we make the field redefinition $\varphi = (x^-)^2 \phi$, then 
\bea
r^2 \partial_+ \varphi - 2 (x^-)^2 \partial_- \varphi - 2 x^- x^i \partial_i \varphi &=& 0\label{C2}
\eea
The action in flat Minkowski space is given by
\bea
S &=& - \frac{1}{2} \int dx^+ dx^- d^4 x \(2 \partial_+ \phi \partial_- \phi + \partial_i \phi \partial_i \phi\)
\eea
After the field redefinition, the action becomes 
\bea
S &=& - \frac{1}{2} \int \frac{dx^+ dx^- d^4 x}{(x^-)^4} \(2 \partial_+ \varphi \partial_- \varphi + \partial_i \varphi \partial_i \varphi\)
\eea
after we drop one boundary term by assuming that the field drops off sufficiently fast at the lightlike infinities $x^+ \rightarrow \pm \infty$. We may use the constraint to substitute
\bea
\partial_+ \varphi &=& \frac{2 t^2}{r^2} \partial_t \varphi+ \frac{2t}{r} \partial_r \varphi
\eea
into this action, where we now put $t = x^-$. Then the action becomes 
\ben
S &=& - \frac{1}{2} \int \frac{dx^+ dt d^4 x}{t^4} \( \(\partial_r \varphi + \frac{2t}{r} \partial_t \varphi\)^2 + \frac{1}{r^2} G^{\alpha\beta} \partial_{\alpha} \varphi \partial_{\beta} \varphi\)\label{S6d}
\een
where $G_{\alpha\beta}$ is the metric on a unit three-sphere. Let us define a five-dimensional field as 
\bea
u(t,x^i) &=& \varphi(x^+ = 0,t,x^i)
\eea
whose action is
\ben
S_{5d} &=& - \frac{1}{2 g_{YM}^2} \int \frac{dt d^4 x}{t^4} \(\(\partial_r u + \frac{2t}{r} \partial_t u\)^2 + \frac{1}{r^2} G^{\alpha\beta} \partial_{\alpha} u \partial_{\beta} u\)\label{S5d}
\een
Here $g_{YM}^2$ is the Yang-Mills coupling constant, which is proportional to the compactification radius of the orbit that is generated by the vector field $V^M$. For this step we thus assume that we have performed a compactification of the six-manifold by making a periodic identification as we travel a certain distance ($2\pi$ times the compactification radius) along the Killing vector field $V^M$. We may for example start out from the hypersurface $x^+ = 0$ at a point $(t,x^i)$ on the hypersurface, and then we travel a certain distance along the integral line of the vector field to a new point $(x'^+,t',x'^i)$. At this new point, the metric is exactly the same as it was where we started at $(0,t,x^i)$ because we were traveing along a Killing direction. We can therefore identify the two points $(0,t,x^i) \sim (x'^+,t',x'^i)$ and that is one way to generate our compactified spacetime manifold. We could pick another hypersurface, but the end result, the compactified manifold, will become the same. 

The justification for our five-dimensional action is as follows. We can recover the field $\varphi(x^+,t,x^i)$ by solving the equation of motion that we derive from $S_{5d}$ for the field $u(x^-,x^i)$ if we also use the constraint equation that determines how the field depends on $x^+$. All information about the dynamics of the zero mode $\varphi$ is thus captured by the field $u(t,x^i)$ and its action $S_{5d}$.

By solving the equation (\ref{eqOc}) we find the Weyl rescaled metric
\ben
ds^2 &=& \frac{1}{(x^-)^2}\(2 dx^+ dx^- + dx^i dx^i\)\label{WT}
\een
with respect to which $V^M$ is a Killing vector. One may also check explicitly that with respect to this metric, $V^M$ as given above, indeed satisfies the Killing vector equation
\bea
\L_V g_{MN} = 4 g_{MN} x^- + g_{NP} \partial_M V^P + g_{MP} \partial_N V^P = 0
\eea
We perform dimensional reduction along $V^M$ by imposing the constraint
\ben
\L_V \varphi &=& 0\label{C2}
\een
where 
\bea
\varphi &=& (x^-)^2 \phi
\eea
Here we notice that $\phi$ acquires a corresponding Weyl factor $(x^-)^2$ under the Weyl transformation of the metric. We may now also understand the appearance of the factor of $\frac{1}{(x^-)^4}$ that appears in the action (\ref{S6d}) as a measure factor that comes from the Weyl rescaled metric $g_{MN} = \frac{1}{(x^-)^2} \eta_{MN}$ as 
\bea
\sqrt{-g} g^{MN} = \frac{1}{(x^-)^4} \sqrt{-\eta} \eta^{MN}
\eea
Let us now look at a different five-dimensional action, 
\bea
S &=& \frac{1}{2g_{YM}^2} \int \frac{dt d^4 x}{t^4} \(\frac{1}{c^2} (\partial_t \varphi)^2 - \(\partial_r \varphi + \frac{2t}{r} \partial_t \varphi\)^2 - \frac{1}{r^2} G^{\alpha\beta} \partial_{\alpha} \varphi \partial_{\beta} \varphi\)
\eea
where we have added a kinetic energy term for the scalar field. Such an action can be understood as an action of a theory that lives on a Lorentzian five-manifold with the metric 
\bea
ds^2 &=& \frac{1}{t^{8/3}} \(- c^2 \(dt - \frac{2 t}{r} dr\)^2 + dr^2 + r^2 d\Omega_3\) 
\eea
Our action in (\ref{S5d}) is recovered from this action by taking the nonrelativistic limit $c \rightarrow \infty$. The metric in both 6d and in 5d have a singularity at $x^- = 0$. This might be interpreted as a lightlike boundary in the 6d theory, or as an initial time slice in the reduced 5d theory. It would be interesting to work out the supersymmetric extension of this action and check that the action is supersymmtric and  preserves the two supercharges $\eps_I = \Gamma_M \eta_I x^M$.

We could also consider the case where we keep one Poincare supersymmetry $\eps_{I0}$ for say $I = 1$, and one special conformal supersymmetry $\eta_I$ for say $I = 2$. In that case there will be a nonvanishing R-rotation $R^I{}_J$ that is Weyl invariant. Then we need to in addition twist all fields that are charged under the R-symmetry by a corresponding term in the constraint equations that we impose under the dimensional reduction. The form of the vector field $V^M$ changes depending on what supersymmetries we want to preserve and that will result in another conformally flat metric where $V^M$ becomes a Killing vector field.

More generally, by following our procedure, we may obtain a supersymmetric Lagrangian in five dimensions that should be closely related to the six-dimensional theory. The five-dimensional theory could provide us with a window into this six-dimensional theory. The ability to preserve some supersymmetry under the dimensional reduction might be useful in order to probe various aspects of the six-dimensional theory for various six-dimensional geometries. For example one could try to use such Lagrangians in five dimensions for supersymmetric localization.

\section*{Acknowledgement}
This work was supported in part by NRF Grant RS-2023-00208011.

\appendix
\section{Lightlike geometry}
We define
\bea
\Omega &=& \nabla_M V^M\cr
\Xi &=& \nabla_M U^M
\eea
We notice that
\bea
\L_V V^P = V^M \nabla_M V^P - \nabla_M V^P V^M = 0
\eea
and 
\bea
\L_V U^P &=& V^M \nabla_M U^P - U^M \nabla_M V^P
\eea
but we will impose the constraint that the Lie bracket on the right-hand side is zero, 
\bea
[V,U]^P = V^M \nabla_M U^P - U^M \nabla_M V^P = 0
\eea
So we have
\bea
\L_V V^P &=& 0\cr
\L_U U^P &=& 0\cr
\L_V U^P &=& 0\cr
\L_U V^P &=& 0
\eea
Furthermore
\bea
\L_V V_P &=& \frac{\Omega}{3} V_P\cr
\L_U U_P &=& \frac{\Xi}{3} U_P
\eea
These results are telling us that $V^M$ and $U^M$ shall be assigned Weyl scaling dimension zero, and $V_M = g_{MN} V^N$ and $U_M = g_{MN} U^N$ will consequently have $\Delta = 2$, which is the scaling dimension of the metric tensor $g_{MN}$. 

Taken together, these relations imply 
\ben
\L_V \N &=& \frac{\Omega}{3} \N\cr
\L_U \N &=& \frac{\Xi}{3} \N\label{NN}
\een
and
\bea
\L_V \(\frac{V_P}{\N}\) &=& 0\cr
\L_U \(\frac{U_P}{\N}\) &=& 0
\eea

\section{Proofs of $V^M \Omega_{MN} = 0$ and $U^M \Omega_{MN} = 0$}\label{Omegas}
We compute
\bea
V^M \Omega_{MN} &=& V^M \nabla_M \(\frac{U_N}{\N}\) - V^M \nabla_N \(\frac{U_M}{\N}\)\cr
&=& - \frac{\nabla_M \N}{\N^2} V^M U_N + \frac{\nabla_N \N}{\N}\cr
&& + \frac{1}{\N} \(V^M \nabla_M U_N - V^M \nabla_N U_M\)\cr
&=& - \L_V N \frac{U_N}{\N^2} + \frac{1}{\N} \(V^M \nabla_M U_N + U^M \nabla_N V_M\)
\eea
Now we rewrite the last term using the conformal Killing vector equation
\bea
\nabla_M V_N + \nabla_N V_M &=& \frac{\Omega}{3} g_{MN}
\eea
We also use that the Lie bracket vanishes $V^M \nabla_M U_N - U^M \nabla_M V_N = 0$. Then we get
\bea
V^M \Omega_{MN} &=& \(- \L_V \N + \frac{\Omega}{3} \N\) \frac{U_N}{\N^2}
\eea
but what stands in the parentesis is zero by the result in (\ref{NN}). Hence $V^M \Omega_{MN} = 0$.  

Next we compute
\bea
U^M \Omega_{MN} &=& \frac{1}{\N} U^M \nabla_M U_N - \frac{1}{\N^2} U_N U^M \nabla_M \N\cr
&=& \L_U \(\frac{U_N}{\N}\)\cr
&=& 0
\eea

\section{Proof of equation (\ref{p})}\label{PP}
From 
\bea
\Gamma^M \eps_I V_M &=& 0
\eea
we get
\bea
0 &=& \Gamma^N \nabla_N \(\Gamma^M \eps_I V_M\)\cr
&=& - 4 \Gamma^M \eta_I V_M + \eps_I \Omega + \Gamma^{MN} \eps_I \nabla_M V_N
\eea
By combining this with (\ref{LVE}) we obtain
\bea
V^M \nabla_M \eps_I &=& \frac{\Omega}{6} \eps_I + 4 \eps_J R^J{}_I
\eea

\section{Proof of equation (\ref{strange})}
By straightforwardly expanding  
\bea
\L_V \eta_I &=& \frac{1}{6} \L_V \(\Gamma^M \nabla_M \eps_I\)
\eea
by commuting $\L_V$ through the Dirac operator and using our formular for $\L_V \eps_I$, we get
\bea
\L_V \eps_I &=& - \frac{\Omega}{12} \eta_I + 8 \eta_J R^J{}_I\cr
&& + \frac{1}{12} \Gamma^M \eps_I \(\frac{\nabla_M \Omega}{6} - R_{MN} V^N - \frac{1}{2} \nabla^N \(\nabla_N V_M - \nabla_M V_N\)\) 
\eea 
The quantity in the parentesis can be expanded further as
\bea
&& \frac{\nabla_M \Omega}{6} - R_{MN} V^N - \frac{1}{2} \nabla^N \(\nabla_N V_M - \nabla_M V_N\) \cr
&=& \frac{\nabla_M \Omega}{6} - \frac{5}{4} R_{MN} V^N + \frac{R}{8} V_M - 10 \bar\eta^I \Gamma_M \eta_I
\eea
Next 
\bea
\Omega &=& 12 \bar\eps^I \eta_I\cr
\nabla_M \eta_I &=& \frac{R}{80} \Gamma_M \eps_I - \frac{1}{8} R_{MN} \Gamma^N \eps_I
\eea
(where the latter equation is derived in Appendix A.2 in \cite{Gustavsson:2018rcc}) gives 
\bea
\frac{\nabla_M \Omega}{12} &=& \frac{R}{80} V_M - \frac{1}{8} R_{MN} V^N - \bar\eta^I \Gamma_M \eta_I
\eea
Then we see that 
\bea
\L_V \eta_I &=& - \frac{\Omega}{12} \eta_I + 8 \eta_J R^J{}_I + \frac{\nabla_M \Omega}{12} \Gamma^M \eps_I
\eea

An easy consistency check of this formula can be obtained by looking at 
\bea
\frac{1}{12} \L_V \Omega &=& (\L_V \bar\eps^I) \eta_I + \bar\eps^I (\L_V \eta_I)
\eea
by inserting our formulas for these Lie derivatives on the right-hand side.

\section{From $(2,0)$ to $(1,0)$ supersymmetry}
The $(2,0)$ superconformal Lagrangian in \cite{Gustavsson:2023zny} is given by $\L = \L_A + \L_{CS} + \L_m$ where 
\bea
\L_A &=& \frac{1}{2 \N} \t{G}^{MN} \t{F}_{MN} + \frac{1}{2\N} K^M K_M\cr
\L_{CS} &=& - \frac{1}{4\N} \eps^{MNPQRS} \omega(A)_{MNP} \Omega_{QR} U_S\cr
\L^{tensor}_m &=& - \frac{1}{2} (D_M \phi^A)^2 - \frac{R}{10} (\phi^A)^2\cr
&& + \frac{i}{2} \bar\psi \Gamma^M D_M \psi + \frac{e}{2} \bar\psi \Gamma_M \Gamma^A [\psi,\phi^A] V^M
\eea
What is important to note is that we here shall assume that $U^M$ and $V^M$ are ordinary lightlike Killing vectors. With that assumption, this Lagrangian is invariant under 
\bea
\delta \phi^A &=& i \bar\eps \Gamma^A \psi\cr
\delta A_M &=& i \bar\eps \Gamma_{MN} \psi V^N\cr
\delta \psi &=& \frac{1}{4} \Gamma^{MNP} \eps \(- \t{G}_{MN} \frac{V_P}{\N} + \t{F}_{MN} \frac{U_P}{\N} - K_M \frac{U_N V_P}{\N^2}\)\cr
&& + \Gamma^M \Gamma^A \eps D_M \phi^A - 4 \Gamma^A \eta \phi^A - \frac{i e}{2} \Gamma_M \Gamma^{AB} \eps [\phi^A,\phi^B] V^M\cr
\delta G_{MN} &=& - \frac{i}{2} D_Q \(\bar\eps\Gamma^Q \Gamma_{MNP} \psi\) U^P + \frac{e}{2} \bar\eps\Gamma_Q \Gamma_{MNP} \Gamma^A [\psi,\phi^A] U^P V^Q
\eea
provided we impose $\L_V \Phi = 0$ on all the fields $\Phi$ as well as assume that $\L_V \eps = 0$. 

Now we break $(2,0)$ down to $(1,0)$ by imposing the Weyl projection
\bea
\Gamma^5 \eps &=& \eps\cr
\bar\eps \Gamma^5 &=& - \bar\eps
\eea
where $\Gamma^5$ refers to the fifth of the five R-direction gamma matrices $\Gamma^A$. 
This is consistent with the other Weyl projection
\bea
\Gamma \eps &=& - \eps
\eea
since $[\Gamma,\Gamma^5] = 0$. We define 
\bea
\Gamma^5 \lambda &=& - \lambda\cr
\bar\lambda \Gamma^5 &=& \bar\lambda
\eea
We define 
\bea
\Gamma^5 \psi &=& \psi
\eea
When we put the Weyl spinor indices, these will be placed as $\eps_I$, $\lambda_I$ and $\psi^A$ respectively, and be rised and lowered by $\eps^{IJ}$ and $\eps_{AB}$ acting from the right. We define 
\bea
q^{AI} &=& \sum_{A'=1}^4 \phi^{A'} (\sigma^{A'})^{AI}
\eea
The original $(2,0)$ R-symmetry is $SO(5)$ but that is now broken down to $SU(2)$ that acts on the Weyl spinor index $I$ and the other $SU(2)$ acts on the Weyl spinor index $A$. But now we may allow for an arbitrary number $k$ of hypers, by letting $A = 1,..., 2k$ and then we have an $Sp(k)$ flavor symmetry rotating those hypers. Since the original $SO(5)$ R-symmetry is now gone, we hope that recycling of index $A$ will not cause any confusion. 

We will now choose 
\bea
V^M &=& \bar\eps^I \Gamma^M \eps_I
\eea
Then it can be shown that \cite{Bak:2024ihe}
\bea
\Gamma_M \eps_I V^M &=& 0
\eea
This results in a new Lagrangian that is given by $\L = \L_A + \L_{CS} + \L_m^{tensor} + \L_m^{hyper}$ where $\L_A$ and $\L_{CS}$ are as before and where
\bea
\L^{tensor}_m &=& - \frac{1}{2} (D_M \phi)^2 - \frac{R}{10} \phi^2\cr
&& + \frac{i}{2} \bar\lambda^I \Gamma^M D_M \lambda_I + \frac{e}{2} \bar\lambda^I \Gamma_M [\lambda_I,\phi] V^M
\eea
and
\bea
\L_m^{hyper} &=& - \frac{1}{4} D_M q^{AI} D_M q_{IA} - \frac{R}{20} q^{AI} q_{IA}\cr
&& + \frac{i}{2} \bar\psi_A \Gamma^M D_M \psi^A - \frac{e}{2} \bar\psi_A \Gamma_M  [\psi^A,\phi] V^M + e \bar\lambda^I \Gamma_M [\psi^A,q_{IA}] V^M
\eea
The supersymmetries become
\bea
\delta \phi &=& - i \bar\eps^I \lambda_I\cr
\delta A_M &=& i\bar\eps^I \Gamma_{MN}\lambda_I V^N\cr
\delta \lambda_I &=& \frac{1}{4} \Gamma^{MNP} \eps \(- \t{G}_{MN} \frac{V_P}{\N} + \t{F}_{MN} \frac{U_P}{\N} - 4 K_M \frac{U_N V_P}{\N^2}\)\cr
&& + \Gamma^M \eps_I D_M\phi + 4 \eta_I \phi\cr
\delta G_{MN} &=& - \frac{i}{2} D_Q \(\bar\eps^I \Gamma^Q \Gamma_{MNP} \lambda_I\) U^P
\eea
combined with 
\bea
\delta q^{AI} &=& 2 i \bar\eps^I \psi^A\cr
\delta \psi^A &=& - \Gamma^M \eps_I D_M q^{AI} - 4 \eta_I q^{AI} 
\eea

\end{document}